# Growth Optimization of MoSi Thin Film and Measurement of Transport Critical Current Density of its Meander Structure

*Shekhar Chandra Pandey,[1, 2, a)] Shilpam Sharma[1], M. K. Chattopadhyay[1,2]*

[1]Free Electron Laser Utilization Laboratory, Raja Ramanna Centre for Advanced Technology, Indore, Madhya Pradesh - 452 013, India

[2]Homi Bhabha National Institute, Training School Complex, Anushakti Nagar, Mumbai 400 094, India

[a)]Corresponding author: shekharpandey@rrcat.gov.in, shekharpandey7579@gmail.com

**Abstract** Amorphous thin film superconductors are promising alternatives for the development of superconducting radiation detectors, especially superconducting nanowire single photon detectors (SNSPDs) and superconducting microwire single photon detectors (SWSPDs), due to their homogeneous nature, ease of deposition, and superconducting parameters comparable to the materials currently being used. A study on the optimization of the growth technology and superconducting transition temperature ($T_C$) of MoSi thin films grown on $SiO_2$-coated Si substrate is reported here. These films have been synthesized by co-sputtering of Mo and Si targets with varying compositions and thicknesses to achieve optimized $T_C$ values close to that of the bulk. $Mo_{80}Si_{20}$ and $Mo_{83}Si_{17}$ compositions of the film, each with a thickness of 17 nm, exhibited the highest $T_C$ of 6.4 K and 5.9 K, respectively. Additionally, a meander structure with a 17 $\mu$m wire width was patterned to estimate the transport critical current density ($J_C$), which was measured to be $1.4 \times 10^9\ A/m^2$ at 4 K. Variation of the $T_C$ with film thickness and deposition pressure has been studied. Electrical resistance as a function of temperature of the film before and after meandering was also studied. These properties are compatible with the fabrication of superconducting nanowire, microwire and wide strip single photon detectors.

## 1. Introduction

Research on amorphous superconductors is drawing attention due to their uniform structure [1, 2] and weak pinning effects, which influence their vortex dynamics [3]. These unique characteristics make them important for both fundamental studies and practical applications [4, 5]. In particular, their potential in superconducting nanowire single-photon detectors (SNSPDs) has become a focal point of recent research. These detectors have set new standards in infrared photon detection, outperforming the traditional technologies like photomultiplier tubes and avalanche photodiodes by offering lower noise, higher count rates, and improved spectral efficiency [6]. Amorphous thin films are desirable for their practical use, especially when the spatial homogeneity of the film is the key requirement. The amorphous nature of the Mo-Si thin film depends on the proportion of silicon (Si) and molybdenum (Mo). The variation of composition leads to a variation in the superconducting properties. It is well known that transition-metal based alloys superconductors exhibit variation in superconducting properties with composition and disorder [2]. It has been reported that the transition metal-based disordered binary alloys, e.g., Nb-Ti, Nb-Zr, Mo-Ti, Ti-V etc, have higher superconducting transition temperature ($T_C$) than their constituent elements [7]. Flux line pinning at the disorder sites is favorable for high dissipationless current-carrying capability. Transition metal-based amorphous superconductors exhibit interesting properties [8] suitable for the fabrication of superconducting nanowire single photon detectors (SNSPDs) [9, 10]. SNSPDs find applications in quantum communication [11], ground-to-space communication [12], laser medicine [13], and quantum computing [14]. Currently, the Nb-based alloy superconductors and their nitrides are widely employed for SNSPD fabrication [15-17]. However, the amorphous superconducting materials present a promising alternative to the Nb-based alloys due to various advantages [18], including their high uniformity, low superconducting energy gap [19], and tunability in the $T_C$. The low superconducting energy gap of the amorphous materials contributes to a higher intrinsic single-photon detection efficiency. Amorphous W-Si has established itself as a highly efficient material for SNSPDs in near-infrared wavelengths, offering performance that matches or even surpasses traditional NbTiN-based detectors [20, 21]. Its uniformity and favorable superconducting properties contribute to high detection efficiency, making it a preferred choice for many photon detection applications. Mo-Si, on the other hand, has emerged as a promising alternative, demonstrating comparable efficiency while offering advantages such as higher operational temperatures and improved timing jitter [22]. These characteristics make Mo-Si particularly attractive for applications where relaxed cooling requirements and enhanced temporal resolution are beneficial.

In the present work, our motivation is to grow Mo-Si at ambient temperature using co-sputtering of Mo and Si. This approach provides precise control over the composition, allowing fine-tuning of its superconducting properties. Optimizing the deposition conditions can enable better material uniformity and tailored transition temperatures, further enhancing the performance of the Mo-Si thin films for SNSPD applications. Conventional SNSPDs typically employ nanowires with widths in the range of 50– 150 nm. However, in recent years, there has been a growing interest in expanding the active area of superconducting detectors to enhance the optical coupling efficiency and simplify the biasing, especially for applications requiring large-area detection or integration with optical systems [17, 23, 24]. By designing the wire to be significantly wider than the optical spot, the active region for photon absorption can be increased significantly. This is easily achieved in the micron size meander and strips. In such wider structures, the light-absorption process reaches the maximum efficiency, making them ideal for applications where a near perfect photon detection efficiency is required. Moreover, broad strips are more favourable for the lithography techniques as the features of this size are easier to define with high reliability, which improves both the speed and success rate of fabrication. This latter point is especially important for fabricating large detector arrays, which is a crucial requirement for the construction of large photonic quantum computing platforms [34]. To address these challenges, micrometer-scale superconducting microwire and wide-strip single photon detectors [35, 36] have emerged as a promising alternative. Accordingly, in the present work, we have deposited the MoSi meander structure in the form of micron-wide wires and have studied their properties.

Here, we have optimized the $T_C$ of the Mo-Si thin films with composition, thickness and deposition pressure. The present work aims to offer an understanding of the behavior of the Mo-Si thin films by examining how their superconducting properties change with varying growth conditions, composition, and thickness. The Mo-Si thin films were deposited using the DC/Pulsed DC magnetron co-sputtering technique, which enables precise control of composition and film uniformity. This technique ensures the consistency of superconducting properties, which is essential for device applications. Similar trends in the dependence of superconducting properties on thickness and composition have also been extensively studied in the MoGe thin films [25], which is considered to be a potential system for nanowire single photon detectors [19]. MoGe has been widely used in the study of the superconductor-insulator transition (SIT), and the variation of $Tc$ with both Ge content and film thickness is also well established [26]. These results call for a detailed study on the iso-structural material MoSi, more specifically, on MoSi thin films deposited under conditions similar to those of the MoGe thin films. Such studies are important to better understand the effects of sputtering parameters on the superconductivity in amorphous thin films in general.

The electrical resistance as a function of temperature [$R(T)$] of the films was measured across different film compositions and thicknesses to find the $T_C$, thus providing insight into how the deposition conditions influence the superconducting phase. Additionally, to determine the transport critical current density ($J_C$), a meander structure was fabricated using a home-built optical lithography setup. The patterning process was carried out using a lift-off technique, allowing for well-defined structures with feature sizes in the micron range. This approach ensures precise control over the geometry, which is crucial for obtaining reliable electrical measurements. The fabricated meander structure was then characterized to analyze its current-carrying capabilities and overall superconducting performance. By studying the transport properties, the influence of structural and compositional variations on the critical current density was assessed. These investigations are essential for optimizing Mo-Si films for superconducting radiation detector and possible quantum technology applications.

## 2. Experimental Details

Thin films of Mo-Si alloys with thickness 5- 17 nm were deposited through co-sputtering technique using a home-built DC magnetron sputtering system. High-purity Mo (99.95%) and Si (99.999%) targets were used for the deposition. The films were deposited in a ultra-high vacuum environment on Si (100) substrates (dimensions: 10 mm × 10 mm × 0.5 mm) coated with a 300 thick thermally grown SiO₂ layer. The composition of the MoSi films was estimated from the thickness ratio of the Mo and Si layers during co-sputtering, as the molar volumes of the elements are proportional to their deposited thicknesses. In our study, films with two different MoSi compositions were prepared by controlling the ratio of deposition rates for Mo and Si during the deposition. To calibrate the deposition rates, we measured the individual thicknesses of Mo and Si films deposited at different sputtering currents using an Alpha-Step, Tencor stylus surface profilometer, and X-ray reflectivity (XRR). We observed a linear relationship between thickness and sputtering current for each material. This calibration allowed us to accurately control the molar volume ratio of Mo and Si, thereby achieving the desired alloy composition. We have used this same method in our previous work, where it was validated using Scanning Electron Microscopy and Energy Dispersive Spectroscopy [27]. Additionally, a similar methodology has also been used and reported in the literature [28]. The deposition rates for Mo and Si, as a function of sputtering time and current was estimated $16.8 \times 10^{-3}$ ($\pm 2 \times 10^{-5}$) Å/(mA·sec) and $15.4 \times 10^{-3}$ ($\pm 8 \times 10^{-5}$) Å/(mA·sec) respectively. Two out of four magnetron guns, arranged in a confocal geometry, were simultaneously powered for the growth of alloy thin films. A load lock chamber was connected to the main deposition chamber to prevent contamination during substrate transfer. Substrates were cleaned ultrasonically in de-ionized water, followed by boiling acetone and ethyl alcohol, and then dried at 100°C before being loaded into the substrate holder. To ensure film homogeneity, the substrates were rotated at 15 RPM in the focal plane of the magnetron guns. The thin films were deposited at room temperature under a background argon gas pressure of 0.2 Pa ($2 \times 10^{-3}$ mbar) to 0.3 Pa ($3 \times 10^{-3}$ mbar), using high-purity argon gas (99.9995%) with inflow controlled by a mass flow controller. Prior to the deposition, the main chamber was evacuated to a pressure of less than $1 \times 10^{-5}$ Pa ($1 \times 10^{-7}$ mbar). Films of different compositions and thicknesses were deposited by varying the sputtering currents and time for the Mo and Si targets.

The structuring and patterning of the thin films for doing $J_C$ measurements were performed using a home-built optical photolithography setup. This setup utilizes a compound microscope equipped with objective lenses of different magnifications to focus a 405 nm UV diode laser onto the photoresist-coated substrate. The focused laser beam is rastered across the surface to define the desired structures. After exposure, the substrate undergoes a development process in which the exposed regions of the photoresist are removed, creating a pattern. The thin film is then deposited over the entire substrate, covering both the patterned and unpatterned regions. Finally, the sample is subjected to a lift-off process, where it is immersed in acetone that dissolves the remaining photoresist, removing the unwanted film and leaving behind the precisely defined structures required for the $J_C$ measurements.

$R(T)$ and low temperature $J_C$ measurements were performed in the temperature range 2-300 K, in a Cryo-free Spectromag CFSM7T-1.5 magneto-optical cryostat system manufactured by Oxford Instruments, UK. The temperature was measured using a Cernox cryogenic temperature sensor, which has an error of $\pm\ 0.01$ K at 4 K. For the measurements of $R(T)$ and current-voltage characteristics (*I-V* characteristics, for finding $J_C$), probes were arranged in the Van der Pauw geometry. To establish the electrical connections, thin copper wires were used, and high-conductivity silver paint was applied. The superconducting transition temperature $T_C$ was extracted from the $R(T)$ curve by first linearly extrapolating the normal-state resistance $R_N$ to zero temperature to find $R_0$, and then defining the the onset temperature of the superconducting transition as the one that corresponds to 90% of $R_0$ and the offset temperature as the one that corresponds to 10% of $R_0$. The transition width $\Delta T_C$ was estimated from the difference between the onset and offset temperatures. A Keithley 6221 current source was used to apply the bias to the structured film. During the measurement of *I-V* characteristics, the current was systematically swept, and the corresponding voltage was measured at various temperatures using a Keithley 2182A nanovoltmeter.

## 3. Results and Discussion

Mo-Si thin films were deposited by varying both the sputtering current and the background argon pressure. For the $Mo_{83}Si_{17}$ composition, co-sputtering of Mo and Si was done using 157 mA and 45 mA currents respectively. Initially, the film was deposited at a pressure of 0.24 Pa ($2.4 \times 10^{-3}$ mbar), resulting in a $T_C$ of 5.9 K. When the same composition was deposited at a higher pressure of 0.41 Pa ($4.1 \times 10^{-3}$ mbar), the $T_C$ decreased to 5 K. In both the cases, the film thickness was measured to be 17 nm ($\pm\ 1$ nm) using a surface profilometer, confirming that the deposition rate remained consistent. This result indicates that increasing the deposition pressure lowers the superconducting transition temperature. The observed reduction in the superconducting transition temperature with increasing deposition pressure can be explained by the reduced kinetic energy of sputtered atoms due to more frequent collisions in the plasma. At higher pressures, the mean free path of the sputtered species decreases, leading to lower-energy ad-atoms arriving at the substrate surface. This limits their surface mobility and can result in a more disordered film structure. Increased disorder and reduced film density are known to suppress superconductivity by enhancing the electron scattering and reducing the electronic coherence. Therefore, the lower $T_C$ at higher deposition pressures may be a consequence of increased structural disorder in the films. Since a higher $T_C$ was observed at the lower pressure of 0.24 Pa ($2.4 \times 10^{-3}$ mbar), we deposited another composition, $Mo_{80}Si_{20}$, while keeping the deposition pressure fixed at this low value. To further optimize the superconducting properties, the sputtering currents were varied. The highest $T_C$ of 6.4 K was achieved when Mo and Si were co-sputtered with 150 mA and 53 mA currents respectively. Figure 1(a) shows the $T_C$ values for the Mo-Si thin films with these two compositions.

After optimizing the composition, we deposited $Mo_{80}Si_{20}$ films with reduced thickness, as SNSPDs typically require thinner films to enhance the hotspot formation. To investigate the superconducting properties at these reduced thicknesses, we deposited films of 10 nm and 5 nm while maintaining a constant deposition pressure of 0.24 Pa ($2.4 \times 10^{-3}$ mbar). Figure 1(b) shows the $T_C$ for these films, which were measured as 5.6 K and 4.8 K for the 10 nm and 5 nm thick films respectively. Table 1 summarizes the superconducting transition temperature ($T_C$ onset and $T_C$ offset), superconducting transition width ($\Delta T_C$), extrapolated resistance at 0 K ($R_0$), and the resistance at 295 K ($R_{295K}$) for all the films. The observed reduction in $T_C$ with decreasing film thickness is due to the quantum size effect, where reduced thickness leads to increased electron scattering and confinement- suppressing the superconductivity [29]. When the film is very thin, the electrons experience more scattering from the surface and defects, which hinders the formation of Cooper pairs. Additionally, as the thickness reduces, the material behaves more like a two-dimensional system, where the superconductivity is weakened by quantum fluctuations [30]. Strain from the substrate or the surrounding environment can further modify the properties of the film, reducing the $T_C$ [31]. Therefore, controlling the thickness is crucial for optimizing the superconducting properties of the thin films for technological applications. Additionally, our thinner films exhibit a higher normal-state resistance, which is consistent with enhanced disorder and reduced carrier mean free path. The variation of $T_C$ with silicon content suggests that there is an optimal Si concentration where the superconducting amorphous Mo-Si phase is the most stable. As the silicon content deviates from this optimum, the $T_C$ decreases, probably due to an increased fraction of the crystalline phase which disrupts the superconducting properties of the amorphous Mo-Si matrix.

For the estimation of transport $J_C$, a meander structure of $Mo_{80}Si_{20}$ thin film having 17 μm wire-width was synthesized with a thickness of 17 nm with the help of the X and Y raster stages of the lithography unit. Figure 2(a) and 2(b) respectively show the images of the meander structure and the same along with contact pads. $Mo_{80}Si_{20}$ was deposited onto the pre-patterned area, and the desired meander structure was obtained by a lift-off process. The contact pads, 35 nm thick, were made using titanium (Ti) in a second lithography step. Although the meander contains sharp bends, the current crowding effects- which are typically significant in nano-scale devices- are expected to be negligible in this case as the wire is relatively wider and this allows the current to redistribute effectively.

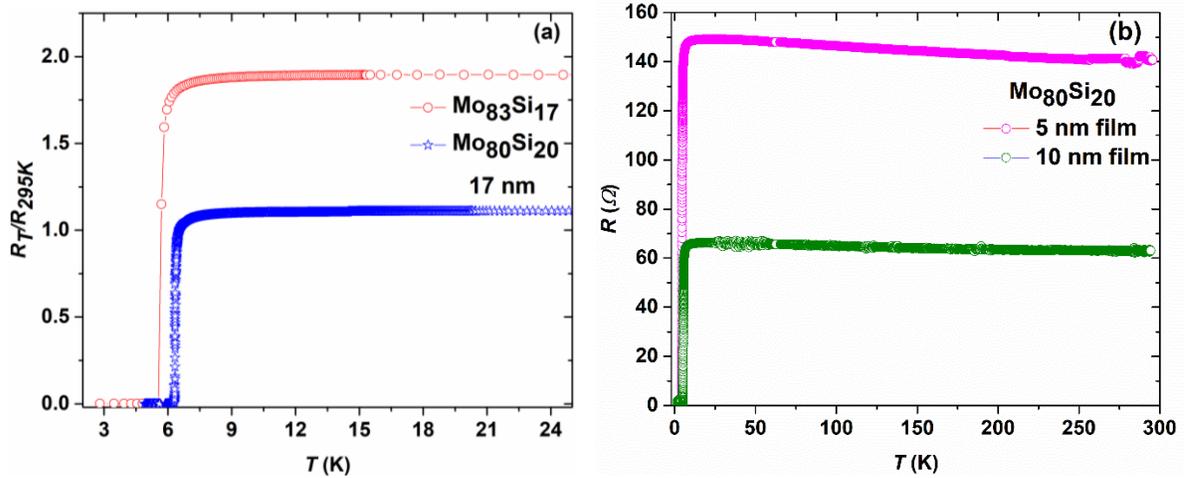

**FIGURE 1. (a)** Normalized resistance as a function of temperature for $Mo_{83}Si_{17}$ and $Mo_{80}Si_{20}$ thin films. Here, $R_{295K}$ denotes the resistance measured at 295 K for each composition individually, and is used to normalize the respective $R(T)$ curves for comparison. The $T_C$ of the $Mo_{80}Si_{20}$ composition is higher than that of $Mo_{83}Si_{17}$. **(b)** $R(T)$ curves for the $Mo_{80}Si_{20}$ thin films having two different thicknesses. The $T_C$ decreases and normal state resistance ($R_N$) increases with decreasing film thickness.

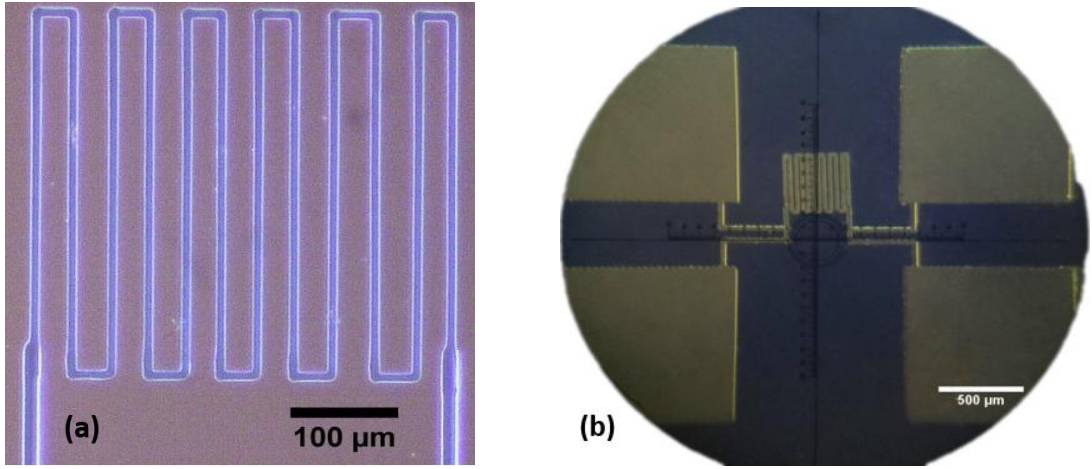

**FIGURE 2. (a)** Meander structure made using optical lithography. The width of each meander wire is 17 micron. **(b)** Meander structure with titanium contact pads.

**Table 1:** Comparison of $T_C$ values of different samples.

| S. No. | Sample | | $R_{295K}$ | $R_0$ | $T_C$ Onset | $T_C$ Offset | Transition width ($\Delta T_C$) |
|---|---|---|---|---|---|---|---|
| | Composition | Thickness (nm) | ($\Omega$) | ($\Omega$) | (K) | (K) | (K) |
| 1. | $Mo_{80}Si_{20}$ | 17 | 45.1 | 52.3 | 6.4 | 6.3 | 0.1 |
| 2. | $Mo_{80}Si_{20}$ | 10 | 63.1 | 65.8 | 5.6 | 5.3 | 0.3 |
| 3. | $Mo_{80}Si_{20}$ | 5 | 140.7 | 147.1 | 4.8 | 4.6 | 0.2 |
| 4. | $Mo_{83}Si_{17}$ | 17 | 2.4 | 4.6 | 5.9 | 5.6 | 0.3 |

Figure 3 (a) shows the comparison of the *R(T)* characteristics of the film both before and after the meandering process. The meander structure exhibits a higher normal state resistance ($R_N$), which is three orders of magnitude greater than that of the film. The inset in Figure 3(a) presents a comparison between the normalized resistance of the film and the meander structure near the $T_C$. The normalized resistance value ($R_T/R_{295K}$) of the film just above the onset of the normal to superconductor transition is almost the same as that of the meander structure. This suggests that the purity and quality of the meander-structured film is comparable to those of the full uniform film. The normal state resistivity of the film was estimated to be $4.7 \times 10^{-6}$ $\Omega \cdot$ m. The *I–V* curves for the meander structure were recorded using a four-probe configuration. Figure 3(b) shows the *I-V* characteristics for the $Mo_{80}Si_{20}$ meander at different temperatures. The critical current is defined as the minimum current value at which the film exhibits a dissipative resistance even below the $T_C$. To determine the $J_C$, the critical value of the current was divided by the area of the meander wire. The transport $J_C$ value for the 17 μm wide and 17 nm thick meander structure of $Mo_{80}Si_{20}$ at 4 K in zero field was estimated as: $Jc = \frac{413 \times 10^{-6}}{(17 \times 10^{-6})(17 \times 10^{-9})} = 1.4 \times 10^9 \text{ A/m}^2$. The $J_C$ value measured in this work at 4 K is comparable to that reported by Banerjee et al. [2] at 3.6 K. However, Lita et al. [32] and Korneeva et al. [33] reported higher $J_C$ values ~$1.3 \times 10^{10}$ A/m² and ~1.1 to $2.5 \times 10^{10}$ A/m² respectively, for micron-wide and nanowire-patterned meander structures, measured at lower temperatures of 250 mK and 1.7 K. As these values were measured at significantly lower temperatures (e.g., 250 mK and 1.7 K), they may only be expected to be higher as compared to the present measurements at 4 K.

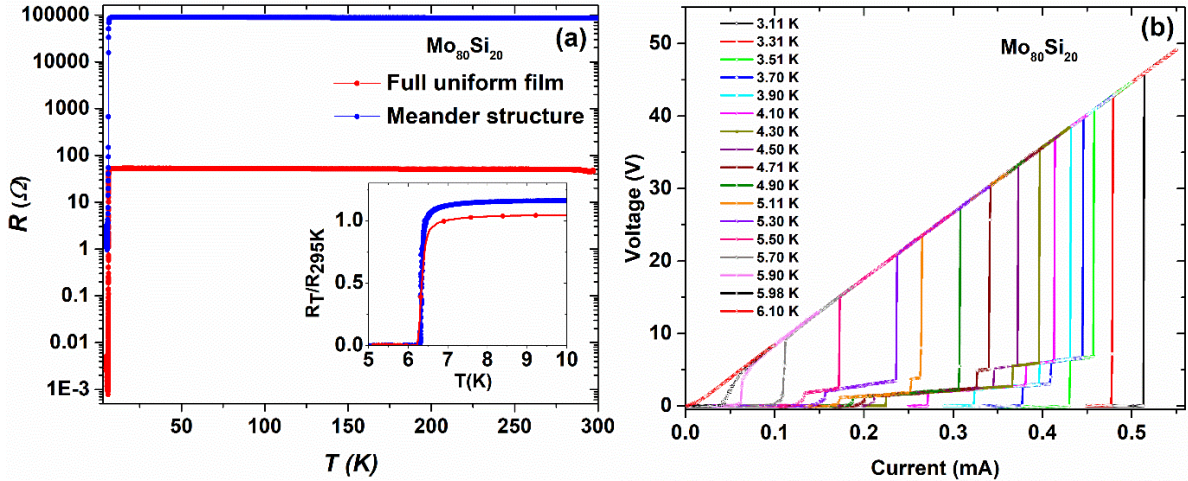

**Figure 3. (a)** Comparison of $R(T)$ curves of $Mo_{80}Si_{20}$ thin film before and after meandering. **(b)** $I$-$V$ characteristics of the $Mo_{80}Si_{20}$ meander at different temperatures.

## 4. Conclusions

In conclusion, MoSi thin films were successfully deposited using the DC/pulsed DC co-sputtering technique and studied in detail using low-temperature transport measurements. Thin films of different compositions and thicknesses were deposited, and the impact of these variations on the $T_C$ and $R_N$ was investigated. The influence of deposition pressure on the $T_C$ and $R_N$ was also explored, revealing that an increase in the deposition pressure results in a decrease in the $T_C$. Additionally, a meander structure with a 17 μm wire-width was patterned, enabling a comparison of the film properties before and after meandering. The $J_C$ of the meander structure was determined by recording *I-V* characteristics at different temperatures. The experimentally obtained transport $J_C$ for the $Mo_{80}Si_{20}$ meander structure at 4 K in zero field is $1.4\times10^9$ A/m$^2$. The study highlights the importance of tuning the Si content to promote the formation of the amorphous phase of MoSi, which is known for its optimal superconducting properties. The tunnability of $T_C$ and the large critical current density demonstrate the suitability of these MoSi thin films for superconducting radiation detector applications.

## Acknowledgement

The authors thank Dr. Raj Mohan S. for his help in using the surface profilometer for measuring the thickness of the films.

## Authors' Contribution

All authors contributed equally to this work.

## Conflict of Interest

The authors declare no conflict of interest.

## Data Availability Statement

The data that support the findings of this study are available from the corresponding author upon reasonable request.

## Keywords

Amorphous superconductors, critical current density, co-sputtering, magnetron sputtering, molybdenum silicide, superconducting microwire single photon detectors, superconducting nanowire single photon detectors, superconducting thin films.